\documentclass[12pt,a4paper]{iopart}

\expandafter\let\csname equation*\endcsname\relax
\expandafter\let\csname endequation*\endcsname\relax
\usepackage{amsmath}
\usepackage{iopams}
\usepackage{graphicx}
\usepackage[breaklinks=true,colorlinks=true,linkcolor=blue,urlcolor=blue,citecolor=blue]{hyperref}
\usepackage{array}
\newcolumntype{L}[1]{>{\raggedright\let\newline\\\arraybackslash\hspace{0pt}}m{#1}}
\newcolumntype{C}[1]{>{\centering\let\newline\\\arraybackslash\hspace{0pt}}m{#1}}
\newcolumntype{R}[1]{>{\raggedleft\let\newline\\\arraybackslash\hspace{0pt}}m{#1}}
\usepackage{subfigure}
\usepackage{graphicx}
\usepackage{dcolumn}
\usepackage{bm}
\usepackage{epstopdf}
\usepackage{epsfig}
\usepackage{url}
\usepackage{textcomp}
\usepackage{stmaryrd}
\usepackage{wasysym}
\usepackage{lipsum,lineno}
\usepackage{enumitem} 

\begin{document}

\title[Variational approximations using Gaussian ansatz]{Variational approximations using Gaussian ansatz, false instability, and its remedy in nonlinear Schr\"{o}dinger lattices}

\author{Rahmi Rusin$^{1,2}$, Rudy Kusdiantara$^{1,3}$, and
	Hadi Susanto$^1$}
 \address{$^1$Department of Mathematical Sciences, University of Essex,
 	Wivenhoe Park, Colchester CO4 3SQ, United Kingdom}
 \address{$^2$Department of Mathematics, Faculty of Mathematics and Natural Sciences, Universitas Indonesia, Ged D Lt.\ 2 FMIPA Kampus UI Depok, 16424, Indonesia}
\address{$^3$Centre of Mathematical Modelling and Simulation,
	Institut Teknologi Bandung, 1st Floor, Labtek III,
	Jl.\ Ganesha No.\ 10, Bandung, 40132, Indonesia}
\ead{\{rrusin,rkusdi,hsusanto\}@essex.ac.uk}

\begin{abstract}
We study the fundamental lattice solitons of the discrete nonlinear Schr\"{o}dinger (DNLS) equation and their stability via a variational method. Using a Gaussian ansatz and comparing the results with numerical computations, we report a novel observation of false instabilities. Comparing with established results and using Vakhitov-Kolokolov criterion, we deduce that the instabilities are due to the ansatz. In the context of using the same type of ansatzs, we provide a remedy by employing multiple Gaussian functions. The results show that the higher the number of Gaussian function used, the better the solution approximation. 
\end{abstract}



\submitto{\jpa}


\section{Introduction}
\label{}

One of the well-known mathematical models that describes many important (discrete) physical systems is the discrete nonlinear Schr\"odinger (DNLS) equation \cite{chris2003discrete,kevrekidis2009discrete}. DNLS has many applications in various fields of physics, both as a discretization of the continuum counterpart or in its own right as a lattice system \cite{kevrekidis2001discrete, anderson1983variational,ablowitz2004discrete}. Two notable applications are as models of coupled optical waveguides \cite{christodoulides2003discretizing} and trapped Bose-Einstein condensates (BECs) in a strong optical lattice \cite{carretero2008nonlinear}. 

As a nonintegrable system, DNLS has no exact solution. Scott and MacNeil \cite{scott1983binding} were the first to study the equation systematically and to report stationary soliton solutions. There is a vast literature on approximate solutions to DNLS equations and their stability \cite{pelinovsky2005stability,pelinovsky2008stability,cuevas2008approximation,weinstein1999excitation}.  Variational Approximation (VA) method is frequently used in this case \cite{malomed1996soliton, kaup2005variational}. VA reduces the infinite dimension of the problem into a finite one by introducing a trial function, usually called ansatz, that involve a finite number of parameters describing the dominant characteristics of the solution. This method is usually used for a conservative system, where the dynamics is governed by the system having a Lagrangian  formulation which leads to a conserved energy. By substituting the ansatz into the Lagrangian, one will obtain an effective Lagrangian that is a function of the variational parameters introduced in the ansatz. Using Euler-Lagrange principle, the critical values of the parameters can be found by solving the associated variational equations.

Studies of discrete solitons in DNLS use an exponential function as the standard VA ansatz, see, e.g., \cite{ malomed1996soliton,cuevas2008approximation,kaup2005variational,chong2012validity} and references therein. The function is chosen because it captures the tail behaviour and at the same time provides an effective Lagrangian with a closed form expression. Its validity is presented in \cite{chong2012validity}, where it is shown that the ansatz captures the dynamics of the original infinite dimensional system for small coupling between lattices. 

On the other hand, when the coupling is strong, i.e., the continuum limit, one uses a sech ansatz that may yield an exact solution \cite{anderson1978variational,malomed2002variational} or a Gaussian function \cite{ilg2016dynamics, anderson1983variational}. A natural question then emerges: can we apply an ansatz that works for all coupling constant? The sech or Gaussian function will yield an intractable effective Lagrangian. However, one may employ numerical approximations to yield a semianalytical method. 

Here, we consider a variational method based on a Gaussian ansatz to the 1-D DNLS equation. Even though the infinite summation in the Lagrangian cannot be evaluated to yield a closed form expression, we may approximate it using only several dominating terms in the strongly discrete case, or using an integral approximation in the large coupling case. 

In this work, we report two important findings: 1) there is an interval of coupling constant in which the on-site (i.e., bond-centered) soliton is unstable, in apparent contradiction with established results  \cite{chong2011variational,cuevas2008approximation,kaup2005variational}, i.e.\ a false instability; 2) by introducing a multiple Gaussian ansatz, we provide a remedy to the false instability. False instabilities of the variational technique perhaps were first reported in \cite{malomed1994vibration}. The stability issue reported in here, however, is novel as it does not belong to the case analysed in \cite{kaup1996variational}, that explained false instabilities to be caused by coupling between modes. Kaup and Vogel \cite{kaup2007quantitative} studied that Gaussian ansatz is not good only when one is interested in soliton interactions. Using the Vakhitov-Kolokolov criterion, we conclude that our instability is due to the shape of the ansatz. 

The paper is organised as follows. In Section \ref{secGauss} we apply VA and present approximate solutions of on-site (i.e., bond-centered) and inter-site (i.e., site-centered) solitons. In Section \ref{secnum} we discuss the presence of false instability and the comparison between the analytical result and the numerical computation. In Section \ref{secmul} we propose a remedy for the false instability by considering a multiple Gausian ansatz and finally, in Section \ref{secconcl} we summarize the work.

\section{Approximation based on Gaussian ansatz} \label{secGauss}

The 1-D DNLS we consider is 
\begin{equation} \label{eq1}
i\dot{u}_n = c\Delta u_n-\omega u_n +|u_n|^2u_n, \quad n \in \mathbb{N}.
\end{equation}
where $\Delta \boxdot_n = \boxdot_{n+1}-2\boxdot_n+\boxdot_{n-1}, u_n$ is a complex-valued function of time $t$ at site $n$, $c$ is the strength of the coupling between adjacent sites which is also called as the dispersion coefficient, and $\omega$ is the propagation constant.

To study discrete solitons of the governing equation using VA, we use the ansatz
\begin{equation}\label{ansatz}
 u_n= Ae^{-a(n-n_0)^2}e^{i\left(\alpha+\beta(n-n_0)+\frac{\gamma}{2}(n-n_0)^2\right)},
\end{equation}
where the set of parameters $X=(x_k)=(A, a, \alpha, \beta, \gamma, n_0)^T$ are functions of $t$. On-site and inter-site solitons correspond to $n_0=0,1/2$, respectively.
The variational equations for the dynamics of the parameters are given by (see equation~\eqref{ve2} in \ref{secVA})
\begin{equation} \label{vareq}
 \text{Re}\left[\sum_{n=-\infty}^{\infty}\left(i\dot{u}_n\right)\left(\frac{\partial u_n^*}{\partial x_j}\right)\right]=\text{Re}\left[\sum_{n=-\infty}^{\infty}\left(-\omega u_n - c\Delta u_n-|u_n|^2u_n\right)\left(\frac{\partial u_n^*}{\partial x_j}\right)\right].
\end{equation}
Explicit computations will yield the system of nonlinear differential equations 
\begin{equation} \label{ODE}
M\dot{X}=F(X),
\end{equation}
where $M=(m_{jk}), j,k=1,2,\cdots, 6,$  is the coefficient matrix and $F(X)=(F_j)$ is a vector of nonlinear functions of the variational parameters. They are given in \ref{secapp1}.

\subsection{Time independent solution and stability}

The equilibrium of \eqref{ODE} corresponds to $\alpha=\beta=\gamma=0$ and the variables $A$ and $a$ satisfying  
\begin{equation} \label{sys}
\eqalign
{A^2\phi_0-(\omega+2c) \varphi_0+ce^{-a}\chi_0 &=0,\\
A^2\phi_2-(\omega+2c) \varphi_2+ce^{-a}\chi_2 &=0,}
\end{equation} 
where $\phi_k,\varphi_k$, and $\chi_k$, $k=0,2$, are given in \ref{secapp1}, evaluated at the equilibrium $X^{(0)}$.  Exact solutions of equation \eqref{sys} can be obtained numerically using a fixed point iteration.

After a solution is obtained, we can discuss its stability. Introducing the linearisation ansatz $X=X^{(0)}+\delta X^{(1)}e^{\lambda t}$, taking a series expansion and keeping only the linear term in $\delta$ yield the generalized eigenvalue problem
\begin{equation} \label{EVP}
\lambda \tilde{M} X^{(1)}= BX^{(1)},
\end{equation}
where
\begin{align*}
	B=(b_{jk}),\quad b_{jk}=\left.\frac{\partial F_j(X)}{\partial x_k}\right|_{X^{(0)}}.
\end{align*}
Explicit expressions of the matrix component $b_{jk}$ are given in \ref{secapp2}. The matrix $\tilde{M}$ is 
\[\tilde{M}=\left.
\begin{pmatrix}
0&0&-A\varphi_0&0&-\frac{1}{2}A\varphi_2&0\\
0&0&A^2\varphi_2&0&\frac{1}{2}A^2\varphi_4&0\\
A \varphi_0&-A^2\varphi_2&0&0&0&0\\
0&0&0&0&0&2aA^2\varphi_2\\
\frac{1}{2}A\varphi_2&-\frac{1}{2}A^2\varphi_4&0&0&0&0\\
0&0&0&-2A^2a\varphi_2&0&0
\end{pmatrix}\right|_{X^{(0)}}.\]
A solution is stable when $\text{Re}\left(\lambda\right)<0$ for all $\lambda$. 

\subsection{Approximate solution and eigenvalue of \eqref{sys} and \eqref{EVP}}

\begin{figure}[tbhp]
	\centering
	\includegraphics[scale=0.6]{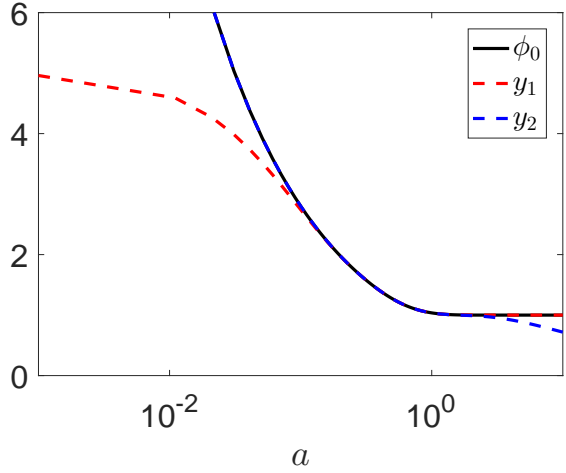}
	\caption{Plot of $\phi_0$ and its approximations $y_1$ and $y_2$.}
	\label{approx}
\end{figure}

To obtain an approximate solution of equation \eqref{sys}, we need to approximate the factors $\phi_k,\varphi_k$, and $\chi_k$, $k=0,2$, analytically. Consider, e.g., 
\begin{equation} \label{fn1}
\phi_0=\sum_{n=-\infty}^{\infty} e^{-4an^2}= \theta_3(0,e^{-4a}),
\end{equation}                                                        
where $\theta_3$ is a theta function defined as \cite{whittaker1996course}
\[\theta_3(u,q)=1+2\sum_{n=1}^{\infty}q^{n^2}\cos(2nu).\] 
For considerably large $a$, the function $e^{-4an^2}$ converges to zero very rapidly as $|n|$ increases. Therefore, we may approximate $\phi_0$ using the first few terms only, e.g.,
$$\phi_0 \approx y_1=1+2e^{-4a_1}+2e^{-16a_1}.$$ On the other hand, for $c\gg\omega$, i.e., $a\to0$, we use the identity
\begin{equation} \label{appr}
\sum_{n=-\infty}^{\infty} f(n)=\int_{-\infty}^{\infty}f(x)\sum_{-\infty}^{\infty}\delta(x-n) dx,
\end{equation}
where $\delta(x)$ is Dirac delta function. Using Fourier series, the Dirac comb can be written as
\[\sum_{-\infty}^{\infty} \delta(x-n) = 1+2 \sum_{k=1}^{\infty} \cos (2k\pi x).\]
Taking only the first harmonic $k=1$, the function $\phi_0$ can be approximated by \[\phi_0 \approx y_2=\frac{\sqrt{\pi}\left(1+2e^{-\frac{\pi^2}{4a}}\right)}{2\sqrt{a}}.\] Figure \ref{approx} shows the comparison between $\phi_0, y_1,$ and $y_2$, where we observe that $y_1$ is indeed good for large $a$, while $y_2$ is good for small $a$. 

After performing the same approximations to the remaining factors, we obtain from \eqref{sys} for the on-site solitons ($n_0=0$) and large $a$ (which holds for $c\ll\omega$)
\begin{equation} \label{sol_onsite}
\begin{aligned}
A& \approx \frac{\sqrt{ (2 c+\omega)(4 e^{-7 a}+e^{-a})-4 e^{-12 a} c-5 e^{-4 a} c-c}}{\sqrt{e^{-15 a} \left(e^{12 a}+4\right)}}, \\
c& \approx \frac{3 \omega}{\kappa+5 \sinh (a)-3 \sinh (3 a)+\cosh (a)+3 \cosh (3 a)-6},
\end{aligned}
\end{equation}
where $\kappa=\frac{e^a \left(20 e^{2 a}-4 e^{4 a}+26 e^{6 a}+9 e^{10 a}+5 e^{12 a}+14\right)}{10 e^{2 a}+10 e^{4 a}+4 e^{6 a}+4 e^{8 a}+e^{14 a}+6}$. At the leading order, when $a \gg 1$, $\kappa \approx 5e^{-a}$ and $c \approx \omega e^{-a}$, which implies that $A$ is well defined. For inter-site solitons ($n_0={1}/{2}$), we obtain the approximation for large $a$ 
\begin{equation} \label{sol_intersite}
\begin{aligned}
A& \approx\frac{16 \sqrt[4]{2} c^{19/8} \sqrt[4]{\omega} \left(2 c^4 \left(5 T+9 \sqrt{\omega}\right)+S_+\right)^{1/2}}{\left(T+\sqrt{\omega}\right)^{17/4} \left(2 c^4 \left(5 T-41 \sqrt{\omega}\right)+9S_-\right)^{1/2}} ,\qquad a \approx \text{arcsinh}\left(\frac{\sqrt{\omega}}{2\sqrt{c}}\right),
\end{aligned}
\end{equation}
where 
$T=\sqrt{4 c+\omega}, \quad S_{\pm}=10 c^3 \omega \left(T \pm 3 \sqrt{\omega}\right)+3 c^2 \left(5 T \omega^2 \pm 9 \omega^{5/2}\right)+c \left(7 T \omega^3 \pm 9 \omega^{7/2}\right)+\omega^4 \left(T \pm \sqrt{\omega}\right)$. The approximate solutions of \eqref{sys} for small $a$ can also be obtained similarly, but we will not present them here. 

Next, we analyse the stability of the solitons in the framework of the VA, i.e., by obtaining their approximate eigenvalue from solving equation \eqref{EVP}. 

For the on-site case, substituting \eqref{sol_onsite} into (\ref{EVP}), we obtain a pair of critical eigenvalues as functions of $A, a$ and $c$
 \begin{equation}\label{eig_onsite}
	\lambda=\pm \frac{\sqrt{c} \left(e^{3 a} \left(c R_1-2 e^{5 a}\omega R_2 \right)+2 A^2 R_3\right)^{1/2}}{e^{19 a/2} \left(16 e^{2 a}+e^{8 a}+18\right)},
	\end{equation}
where
\[
\begin{aligned}
	R_1&=-576 e^{2 a}-480 e^{5 a}-1984 e^{6 a}+1152 e^{7 a}-1608 e^{8 a}-908 e^{10 a}+1688 e^{11 a}+88 e^{12 a}\\
	&\quad+3552 e^{13 a} +1572 e^{14 a}+704 e^{15 a}-780 e^{16 a}-72 e^{17 a}-316 e^{18 a}-652 e^{19 a}+28 e^{20 a}\\
	&\quad-1504 e^{21 a}-28 e^{22 a}-648 e^{23 a}+532 e^{24 a}-104 e^{25 a}+189 e^{26 a}-100 e^{27 a}+24 e^{28 a}\\
	&\quad-28 e^{29 a}+30 e^{30 a}-4 e^{31 a}-4 e^{32 a}-4 e^{33 a}+e^{34 a}+240\\
	R_2&=-288 e^{2 a}-422 e^{6 a}-888 e^{8 a}-176 e^{10 a}+18 e^{12 a}+163 e^{14 a}+376 e^{16 a}+162 e^{18 a} \\
	&\quad+26 e^{20 a}+25 e^{22 a}+7 e^{24 a}+e^{26 a}+e^{28 a}+120\\
	R_3&=-288 e^{2 a}-1082 e^{6 a}-1824 e^{8 a}-640 e^{10 a}+104 e^{12 a}+1230 e^{14 a}+684 e^{16 a}+860 e^{18 a}\\
	&\quad-48 e^{20 a}-221 e^{22 a}+200 e^{24 a}-42 e^{26 a}+24 e^{28 a}+43 e^{30 a}+5 e^{34 a}+120
	\end{aligned}\]
The other eigenvalues are zero or purely imaginary.

For the inter-site case, evaluating the generalized eigenvalue problem \eqref{EVP} at the time independent solution \eqref{sol_intersite} yields a pair of critical eigenvalues   
please replace with the following expressions: 

\begin{equation} \label{eig_intersite}
\lambda= \pm
\frac{2 \sqrt{L_1^8+2 L_1^6-3} \sqrt{c^2L_2+cL_3}}{L_1^4 \left(L_1^4+9\right) \sqrt{L_1^8+9} \sqrt{\ln (L_1)}},
\end{equation}
where \[\begin{aligned}L_1&=\sqrt{\frac{1}{4 c}+1}+\frac{1}{2 \sqrt{c}}\\
L_2&=8 L_1^6 \left(L_1^2-1\right)^2 \left(L_1^4-1\right)-6 \left(2 L_1^6-3 L_1^4+1\right) \left(L_1^8+9\right) \ln (L_1)\\
L_3&=L_1^4 \left(L_1^4+9\right) \left(L_1^8+9\right) \omega \ln (L_1)-8 L_1^8 \left(L_1^4-1\right) \omega.\end{aligned}\] The other eigenvalues are also zero or purely imaginary.

\begin{figure*} [h]
	\centering
	\subfigure[]{\includegraphics[scale=0.33]{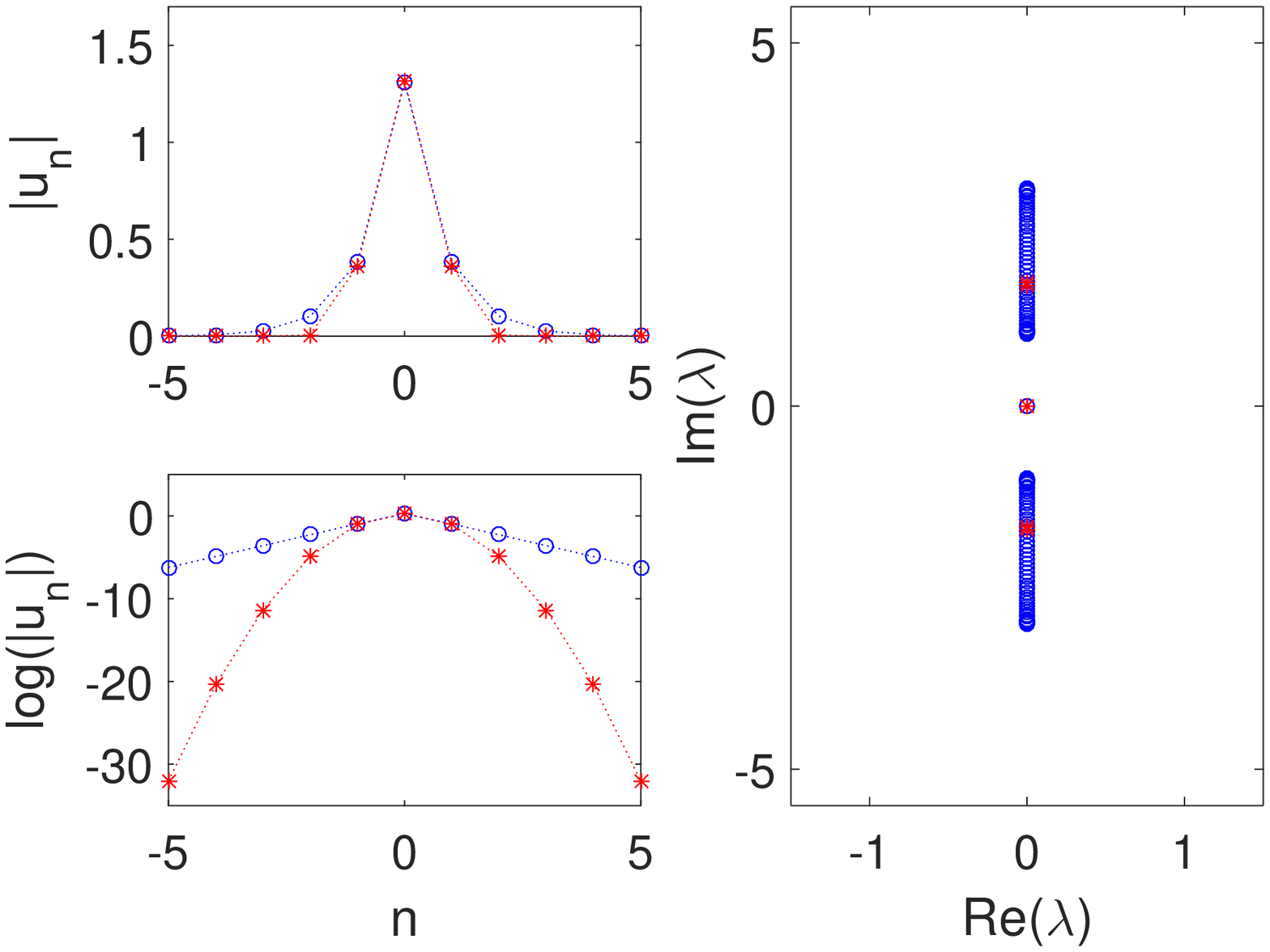}\label{subfig:onsite_N_1_c_0_5}}
	\subfigure[]{\includegraphics[scale=0.33]{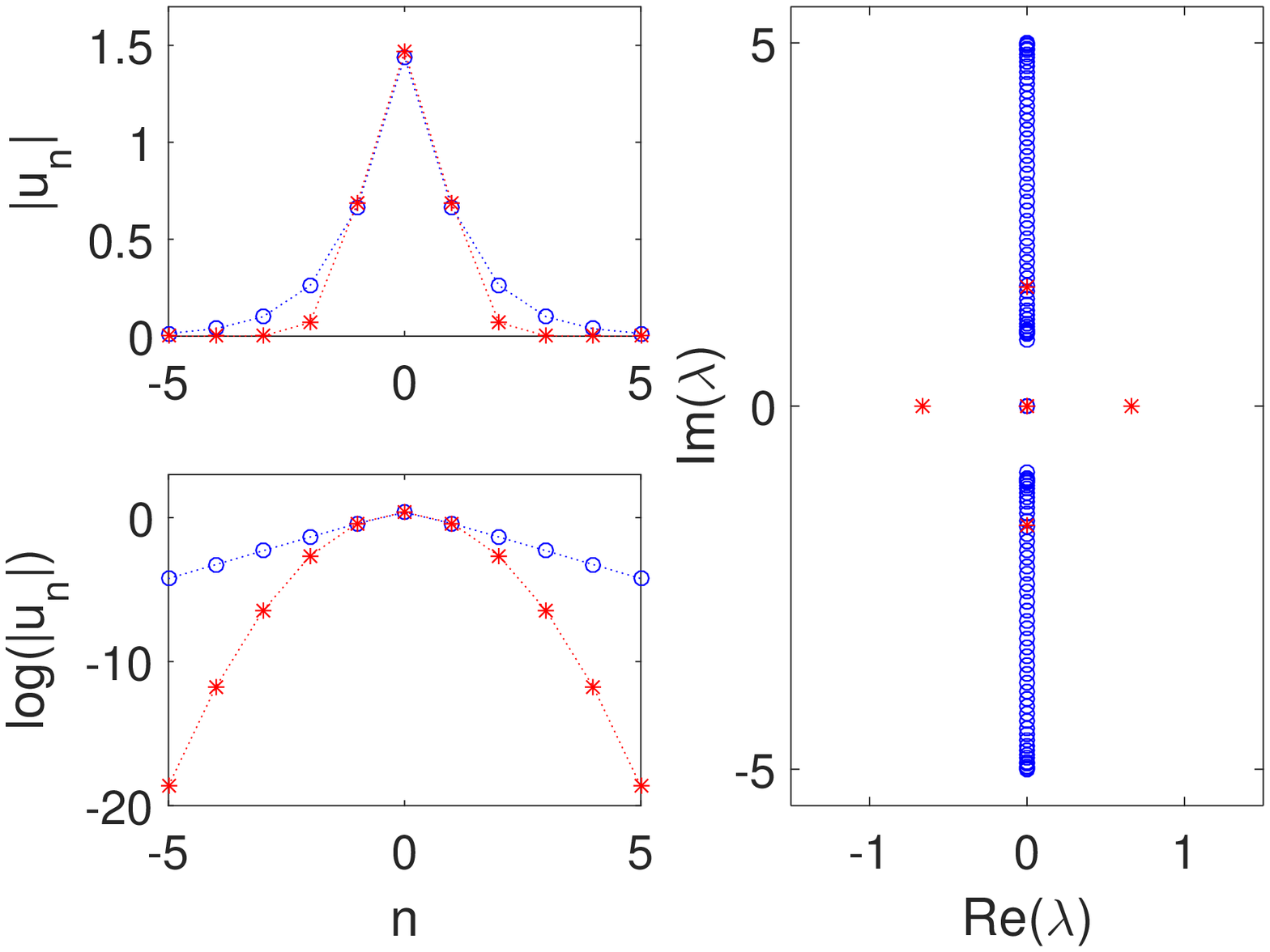}\label{subfig:onsite_N_1_c_1}}
	\subfigure[]{\includegraphics[scale=0.33]{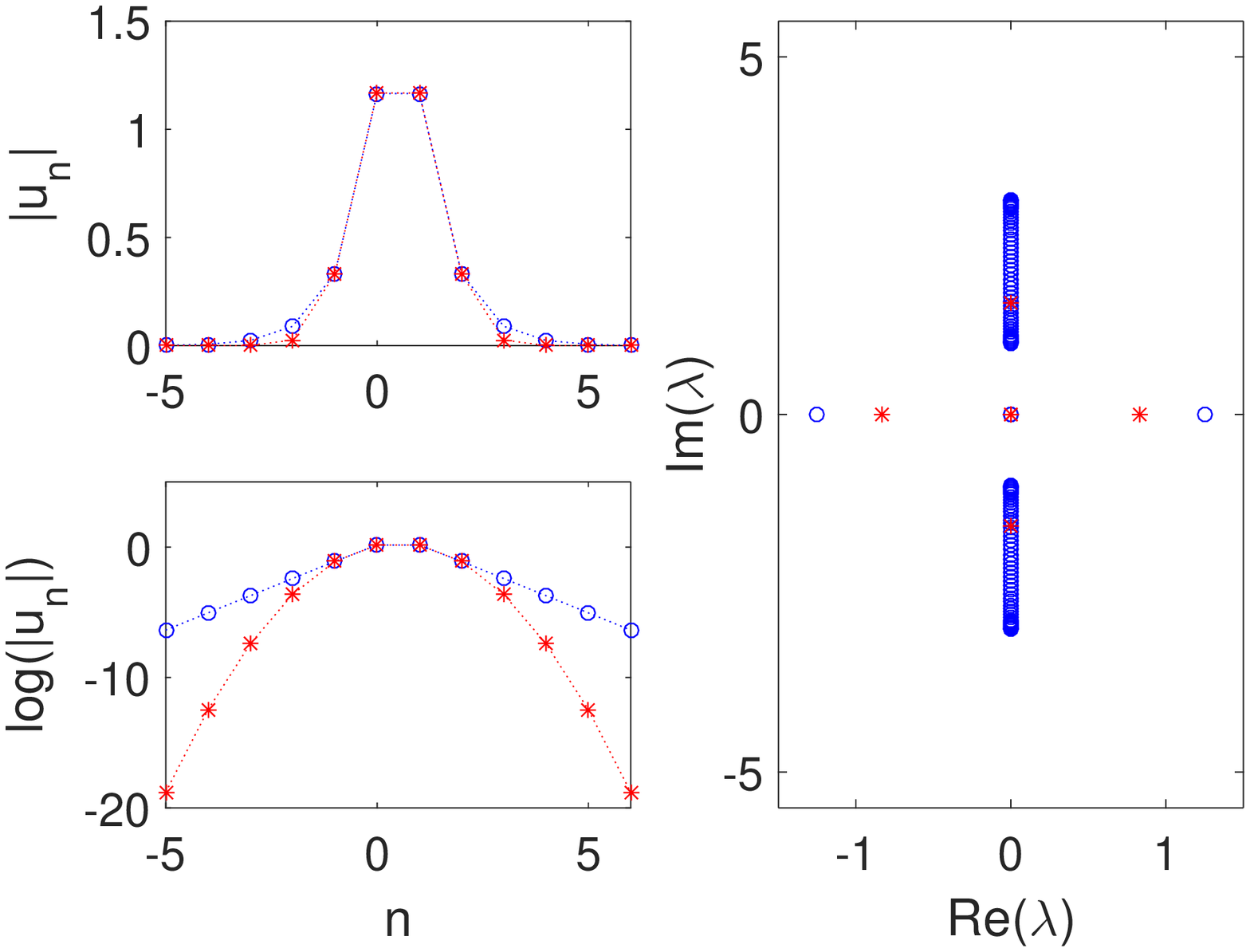}\label{subfig:intersite_N_1_c_0_5}}
	\subfigure[]{\includegraphics[scale=0.33]{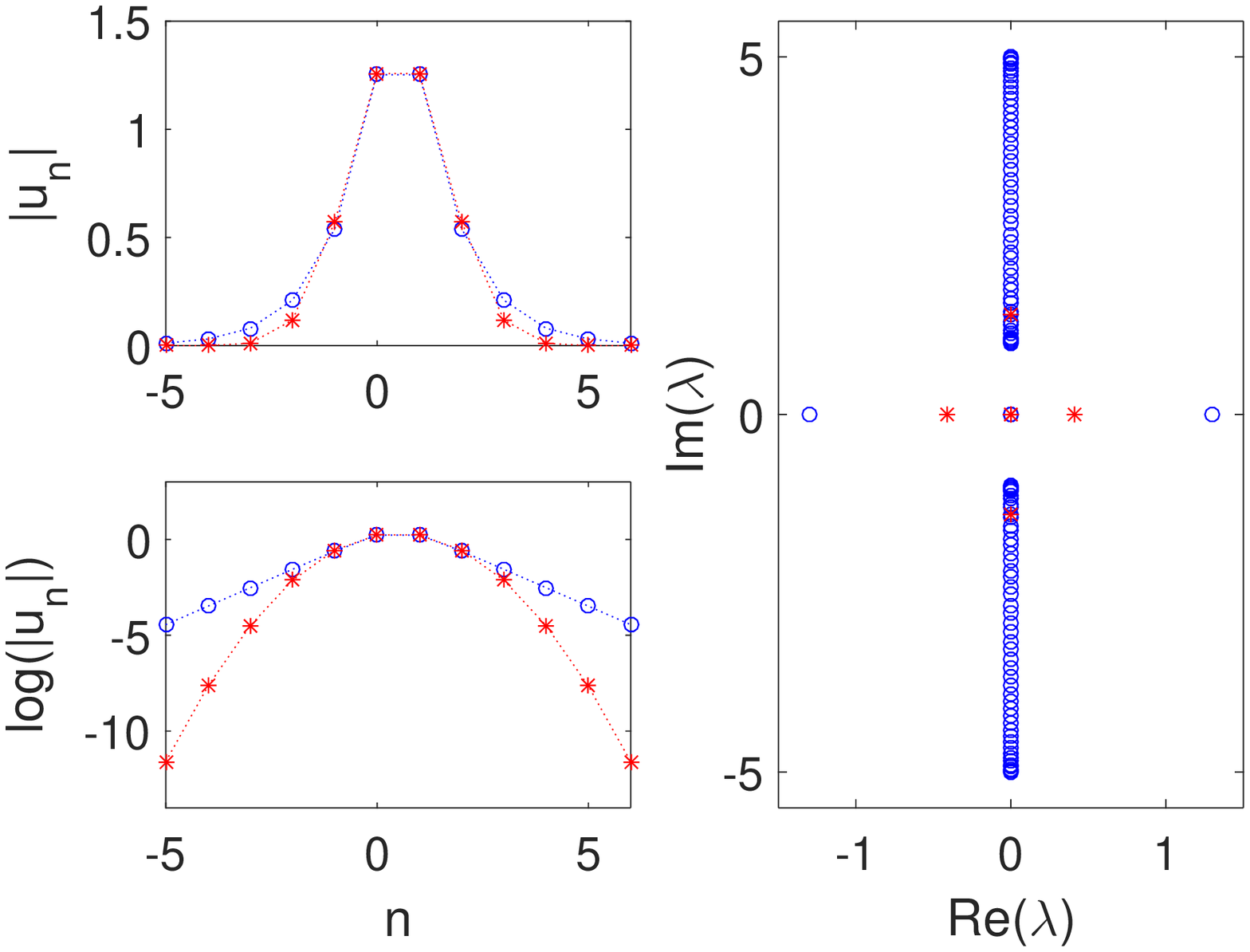}\label{subfig:intersite_N_1_c_1}}
	\caption{Comparison between the numerically obtained on-site (a,b)  and inter-site (c,d) solutions of \eqref{eq1} and Gaussian VA \eqref{ansatz} and \eqref{sys} for $\omega=1$ and (a, c) $c=0.5$, (b, d) $c=1$. Corresponding spectrum of the solution profile is in the right panel of each figure. Blue circle-dashed and red star-dashed lines indicate numerical and Gaussian VA results, respectively. Observe the false unstable spectrum in panel (b).
	}
	\label{fig:onsite_N_1}
\end{figure*}

\section{Numerical comparison and false instability} \label{secnum}

\begin{figure}[h]
	\centering
	\subfigure[]{\includegraphics[scale=0.33]{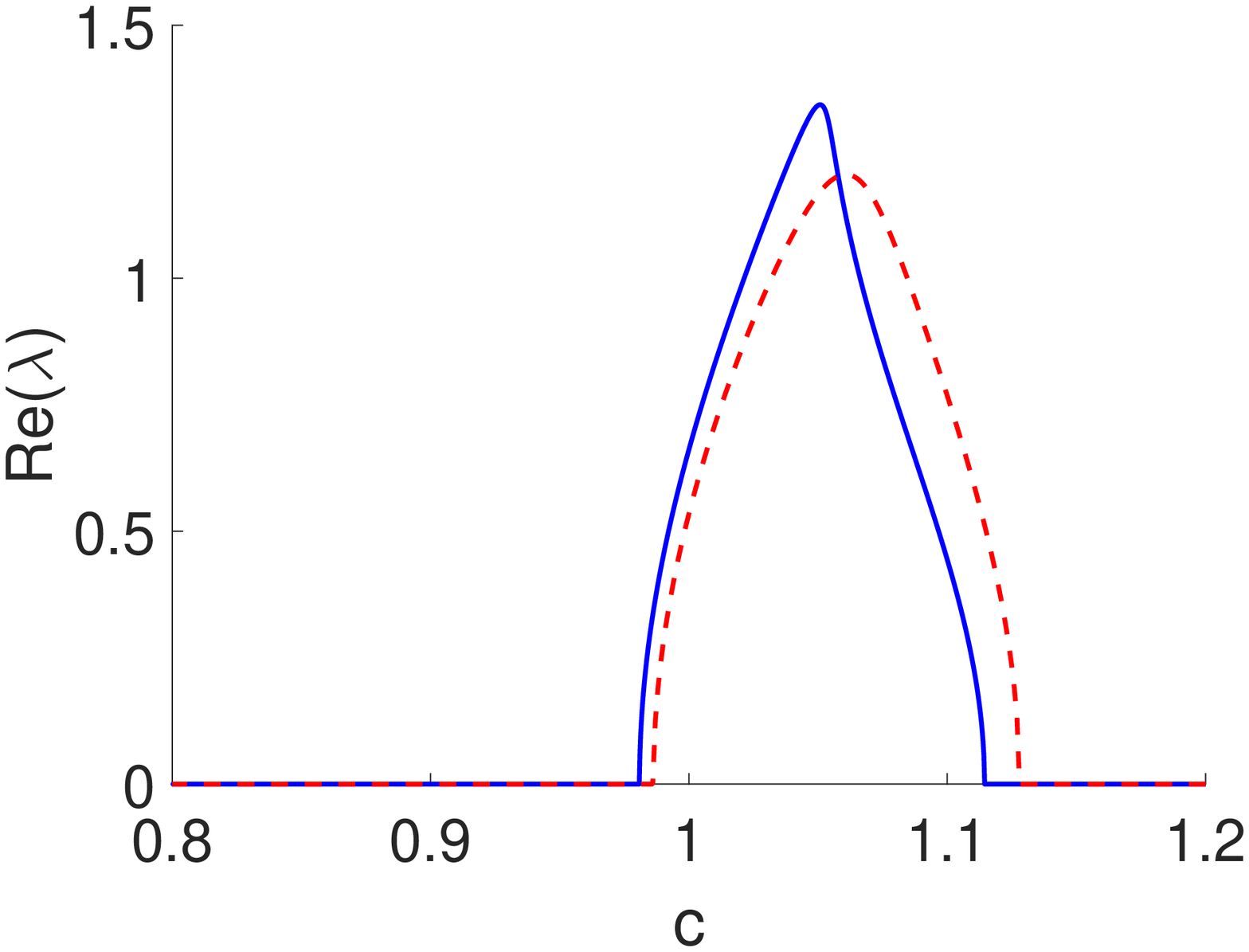}\label{subfig:onsite_stability_real}}
	\subfigure[]{\includegraphics[scale=0.33]{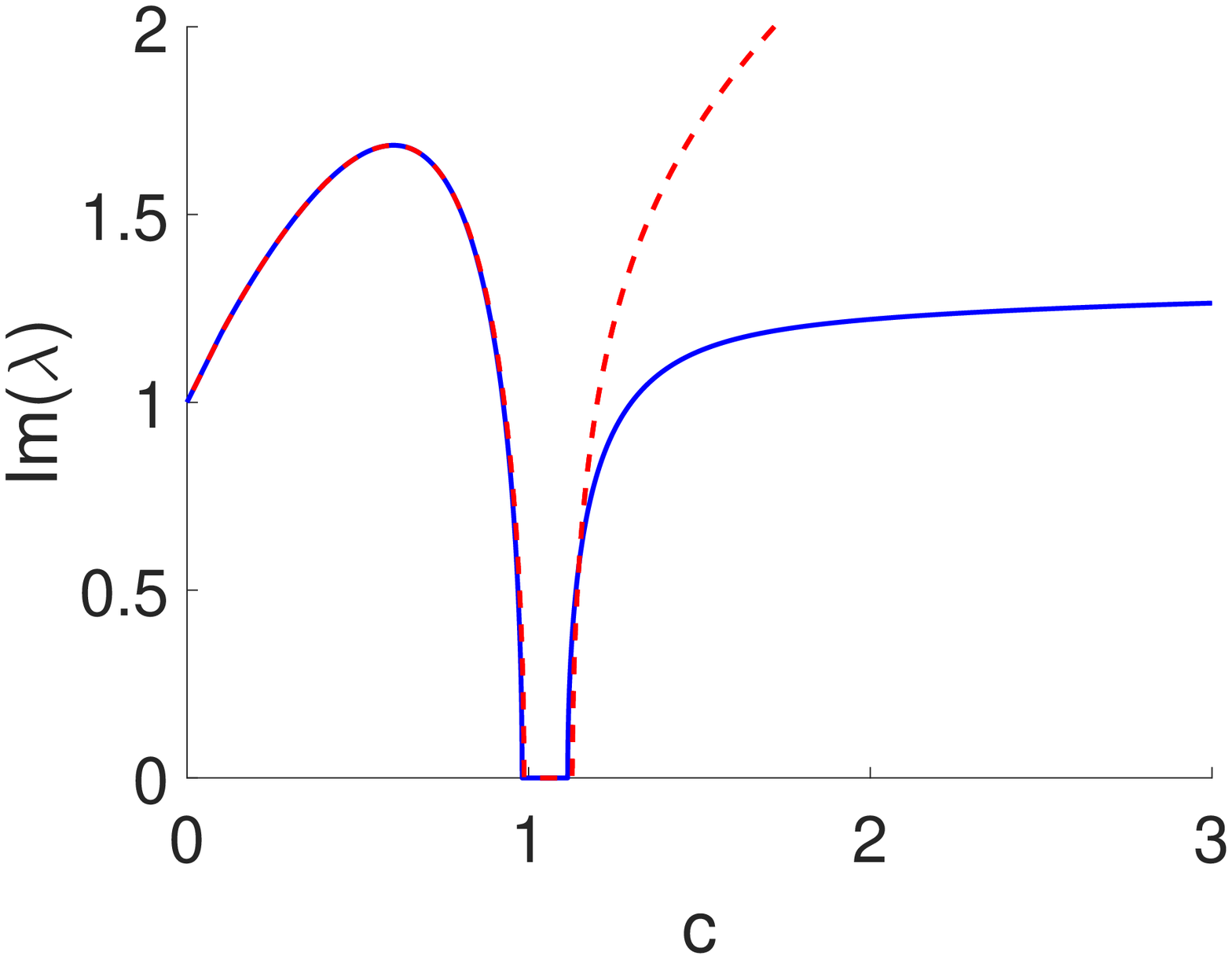}\label{subfig:onsite_stability_imag}}
	\subfigure[]{\includegraphics[scale=0.33]{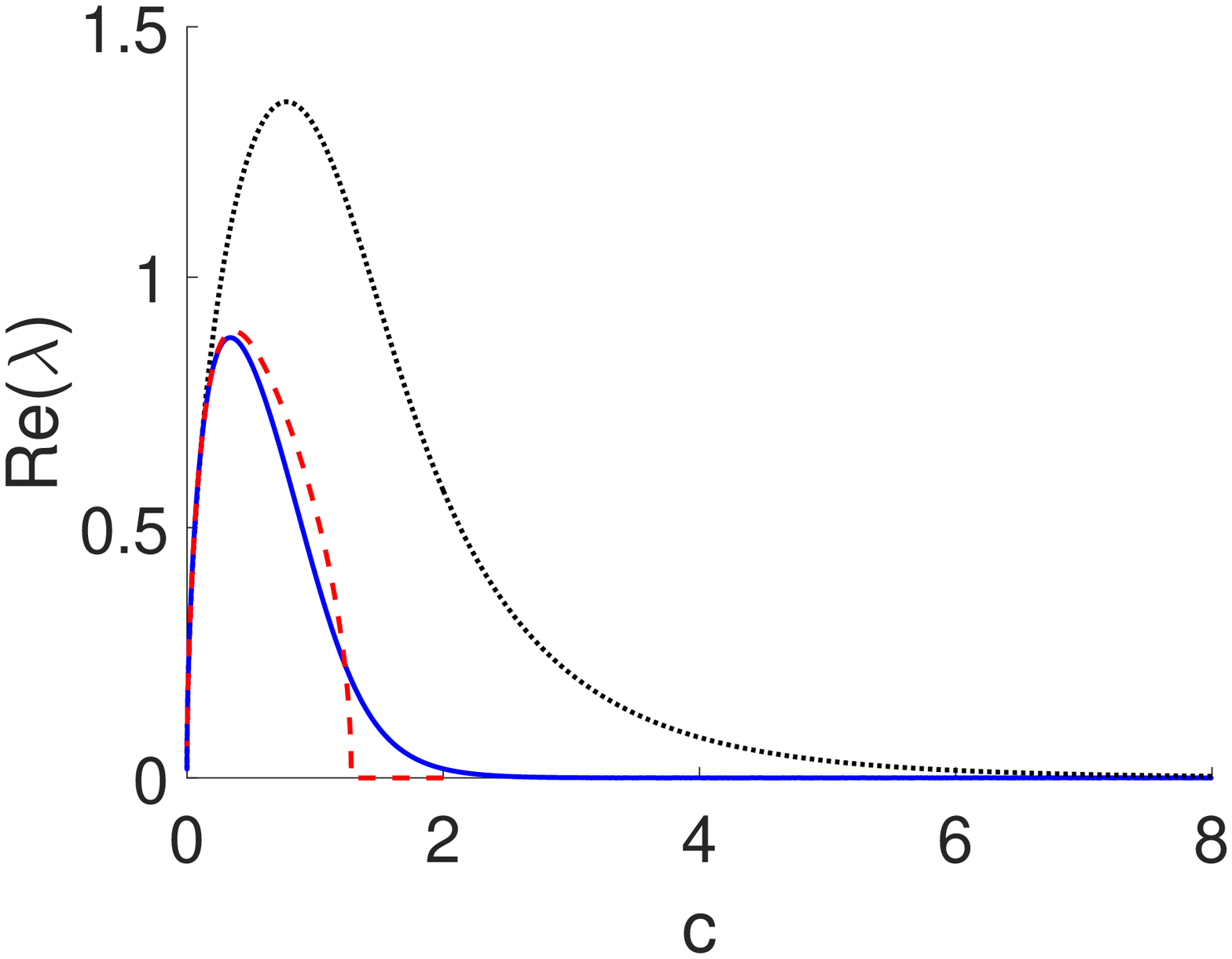}\label{subfig:intersite_stability_real}}
	\subfigure[]{\includegraphics[scale=0.33]{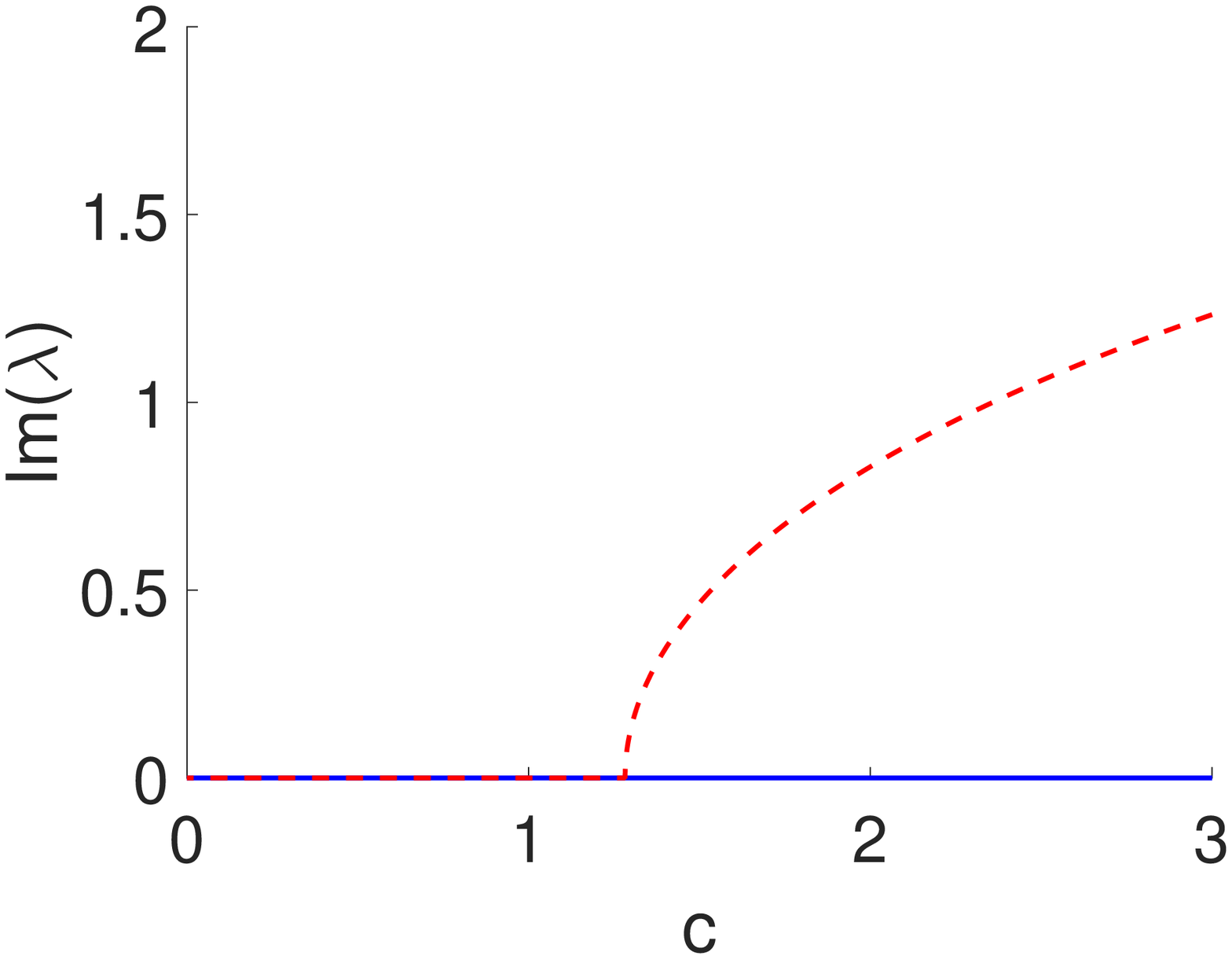}\label{subfig:intersite_stability_imag}}
	\caption{
		Critical eigenvalues calculated using VA of on-site solitons (a,b) and inter-site solitons (c,d) for varying $c$. Red dashed lines in panels (a,b) are equation \eqref{eig_onsite} with $A$ and $a$ using \eqref{sol_onsite}, while those in (c,d) are from equation \eqref{eig_intersite}. Blue solid lines are obtained from solving \eqref{EVP} numerically. In panel (c), the black dotted line shows the critical eigenvalue obtained from solving \eqref{evps}.
			}
	\label{fig:onsite_intersite_stability}
\end{figure}

To check the validity of the VA, we calculate time independent discrete solitons of \eqref{eq1} by solving the system numerically. We used Newton's iteration method. After a discrete soliton, let us say $v_n$, is obtained, we determine its stability by solving the eigenvalue problem using
a linearization ansatz $u_n(t)=v_n+\delta \epsilon_n(t)$,  $\delta \ll 1$. By writing $ \epsilon_n=\left(\eta_n +i\xi_n\right) e^{\lambda t}$, we obtain an eigenvalue problem
\begin{equation}{\label{evps}}
\begin{pmatrix}
0 & -\omega + c\Delta +v_n^2 \\
\omega - c \Delta-3v_n^2 & 0
\end{pmatrix}
\begin{pmatrix}
\eta_n\\
\xi_n
\end{pmatrix}= \lambda
\begin{pmatrix}
\eta_n\\
\xi_n
\end{pmatrix}.
\end{equation}
In the following, unless mentioned otherwise, we set $\omega=1$.

Figures \ref{subfig:onsite_N_1_c_0_5} and \ref{subfig:onsite_N_1_c_1} show the profiles of the time independent on-site soliton and its corresponding spectrum for $c=0.5$ and $c=1$, respectively. The profiles and spectrum of inter-site solitons are shown in Figs.\ \ref{subfig:intersite_N_1_c_0_5} and \ref{subfig:intersite_N_1_c_1}. Results from the original discrete equation \eqref{eq1} and \eqref{EVP} are shown in blue circle-dashed lines. We also have solved equation \eqref{sys} and the eigenvalue problem \eqref{EVP} numerically. Substituting the results into the ansatz \eqref{ansatz}, we plot the VA in Fig.\ \ref{fig:onsite_N_1} in red star-dashed lines. As expected, one can observe that the ansatz is not good in capturing the soliton tails. We have also plotted the eigenvalues from the VA \eqref{EVP}. 

In general, VA captures the qualitative stability of the discrete solitons, i.e., on-site and inter-site solitons are stable and unstable, respectively, due to the presence of an eigenvalue with positive real part. However, it is important to note that we found an unexpected result where according to VA the on-site soliton in Fig. \ref{subfig:onsite_N_1_c_1} is unstable. This finding is in contrast with the established result, that on-site soliton is always stable for any coupling constant.  We therefore observe a false instability. In the following, we will study the emergence of this unstable eigenvalue from our VA.  

In Fig.\ \ref{fig:onsite_intersite_stability}, we plot the eigenvalues of \eqref{EVP} obtained numerically for varying coupling constant $c$. The results are shown in blue solid lines. In Fig.\ \ref{subfig:onsite_stability_real} and  \ref{subfig:onsite_stability_imag} we plot the real and the imaginary parts of the eigenvalues, respectively. From the figures, we conclude that one of the eigenvalues move along the imaginary axis towards the origin and then bifurcates into the real axis, creating a false instability. The unstable eigenvalue exists within a finite interval of coupling constant $c$. In Fig.\ \ref{subfig:onsite_stability_real}, we also plot our analytical approximation \eqref{eig_onsite} as the red dashed line, where good agreement is obtained.

Similarly for the inter-site case, we plot in Figs.~\ref{subfig:intersite_stability_real} and \ref{subfig:intersite_stability_imag} the real and imaginary parts of the eigenvalues obtained from solving \eqref{EVP} numerically as blue solid lines. We also display our approximation \eqref{eig_intersite} as red dashed line, where we again obtain good agreement. As a comparison, we also plot as black dotted curve the critical eigenvalue of inter-site solitons obtained from solving the eigenvalue problem from the original system \eqref{evps}.

\begin{figure}[tbhp]
	\centering
	\includegraphics[scale=0.53]{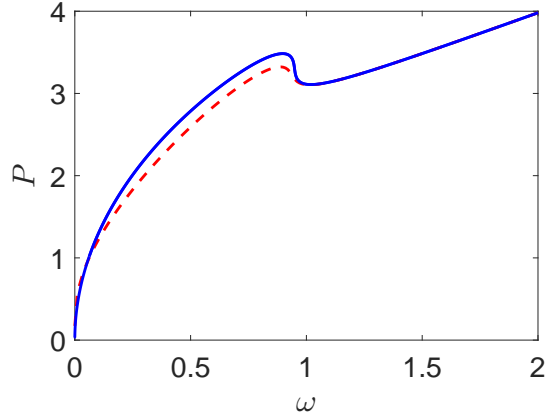}\label{subfig:power_c_1_NN_1}
	\caption{Power of on-site soliton approximated by Gaussian ansatz, as  a function of $\omega$ for $c=1$. Red dashed and blue solid lines correspond to the soliton amplitude $A$ and width $a$ computed in \eqref{sol_onsite} and from \eqref{sys}, respectively. }
	\label{fig:power_onsite_c_1_NN_1}
\end{figure}

The false instability of on-site solitons is believed to be caused by the shape of the ansatz (i.e., the tail error). To show this, we consider the dependence of the soliton power defined as 
\begin{equation}
P(\omega)=\sum_{n=-\infty}^{\infty} |u_n|^2,
\end{equation} 
on the propagation constant $\omega$. Using Vakhitov-Kolokolov criterion \cite{vakhitov1973stationary}, the soliton is unstable when $\frac{dP}{d\omega}<0$. Figure \ref{fig:power_onsite_c_1_NN_1} shows $P$ for varying $\omega$ when $c=1$ for the on-site solitons based on the VA ansatz \eqref{ansatz} and \eqref{sys}. Indeed we obtain a negative slope about $\omega=1$. This confirms our finding that the false instability is due to the Gaussian ansatz \eqref{ansatz}. 

\section{Multiple Gaussian ansatz} \label{secmul}

We will solve the DNLS \eqref{eq1} using VA as in Section \ref{secGauss}, but now using an ansatz containing multiple Gaussian functions,
\begin{equation} 
u_n=\sum_{j=1}^{N} A_je^{-a_j(n-n_0)^2}e^{i\left(\alpha_j+\beta_j(n-n_0)+\frac{\gamma_j}{2}(n-n_0)^2\right)}.
\label{ansatz2}
\end{equation}
We have $5N+1$ parameters, i.e., $A_j,a_j,\alpha_j, \beta_j, \gamma_j, j=1,2,\dots, N$ and $n_0$, being functions of $t$. We will show that it gives a remedy to the false instability reported above.  The idea of using several Gaussian functions in concert here has been proposed and used before in the context of spatially continuous linear or nonlinear Schr\"odinger equations \cite{heller1975time,heller1976time,heller1981frozen,ilg2016dynamics}. However, applying the idea in the context of spatially discrete equations is novel. As the calculation is cumbersome, in the following we only present the results.

\begin{figure*}[h]
	\centering
	\subfigure[]{\includegraphics[scale=0.33]{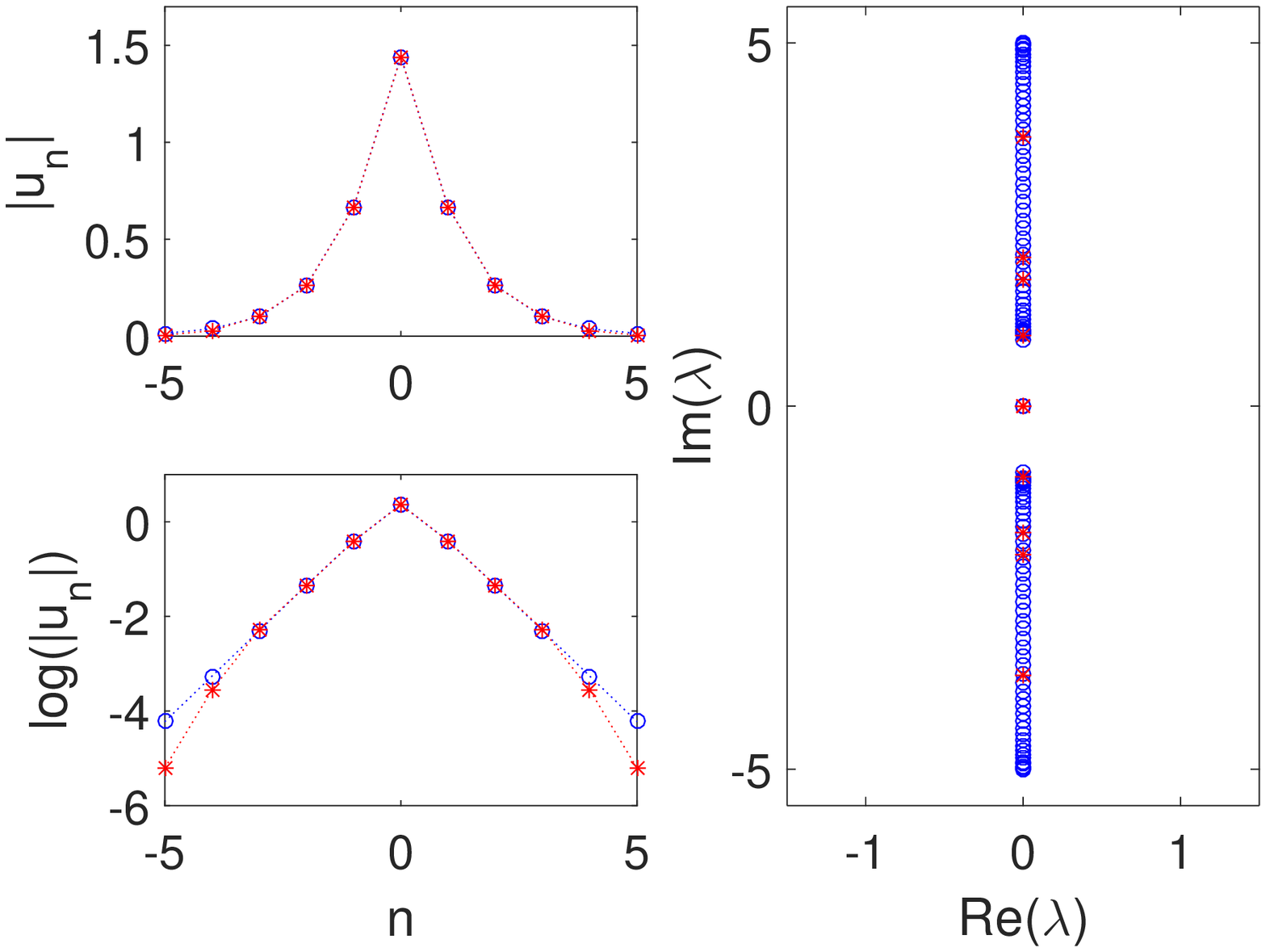}\label{subfig:onsite_N_2_c_1}}
	\subfigure[]{\includegraphics[scale=0.33]{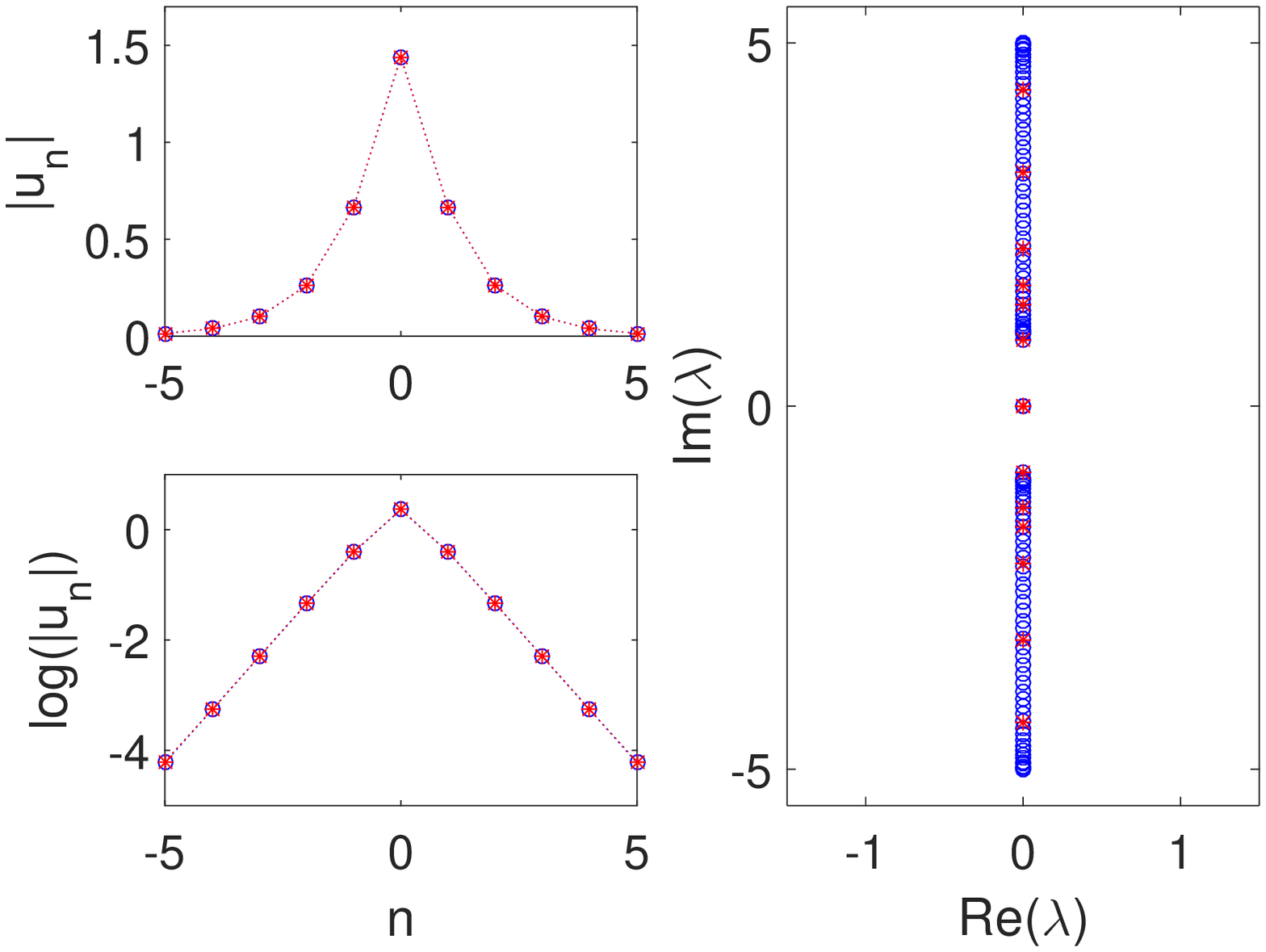}\label{subfig:onsite_N_3_c_1}}
	\subfigure[]{\includegraphics[scale=0.33]{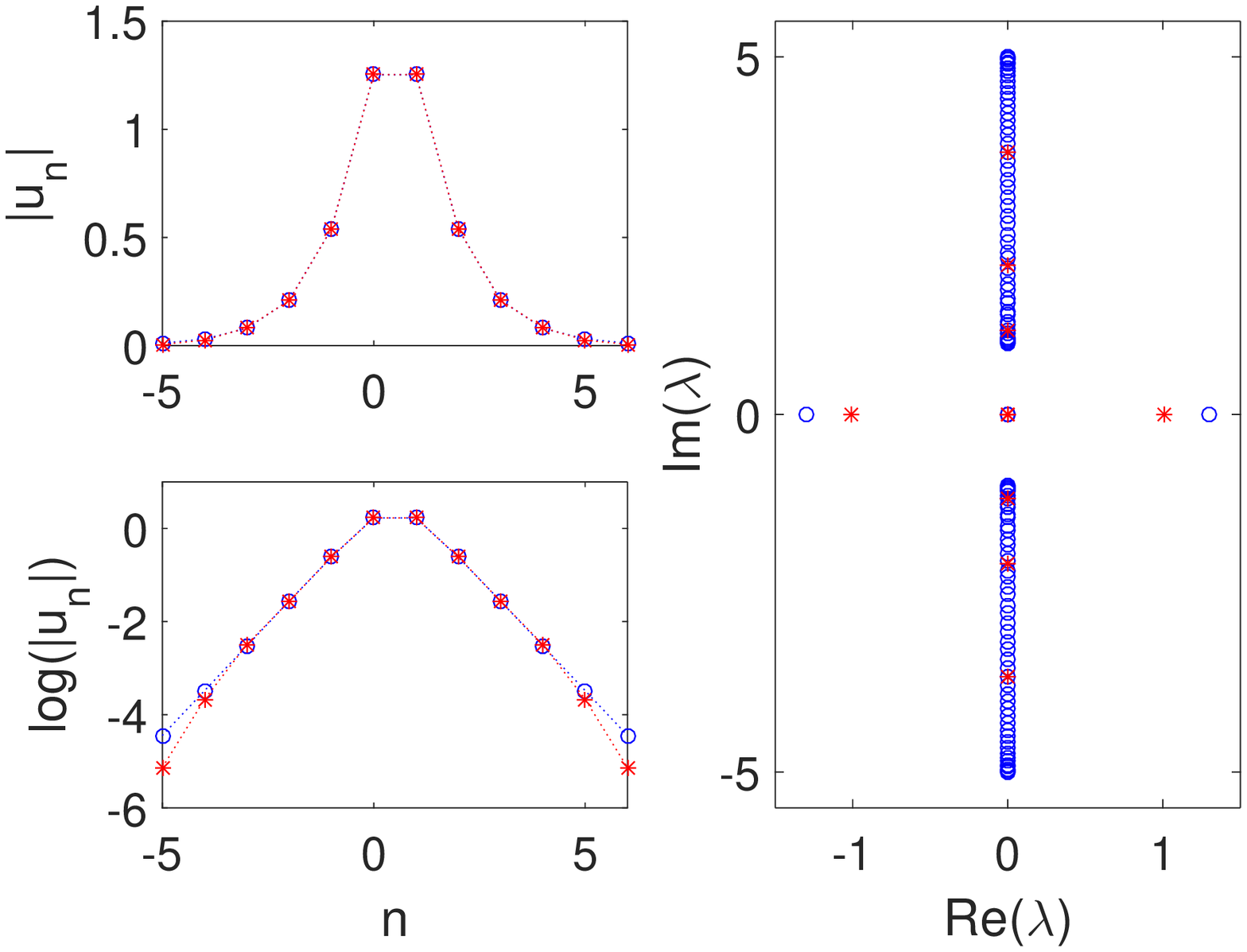}\label{subfig:intersite_N_2_c_1}}
	\subfigure[]{\includegraphics[scale=0.33]{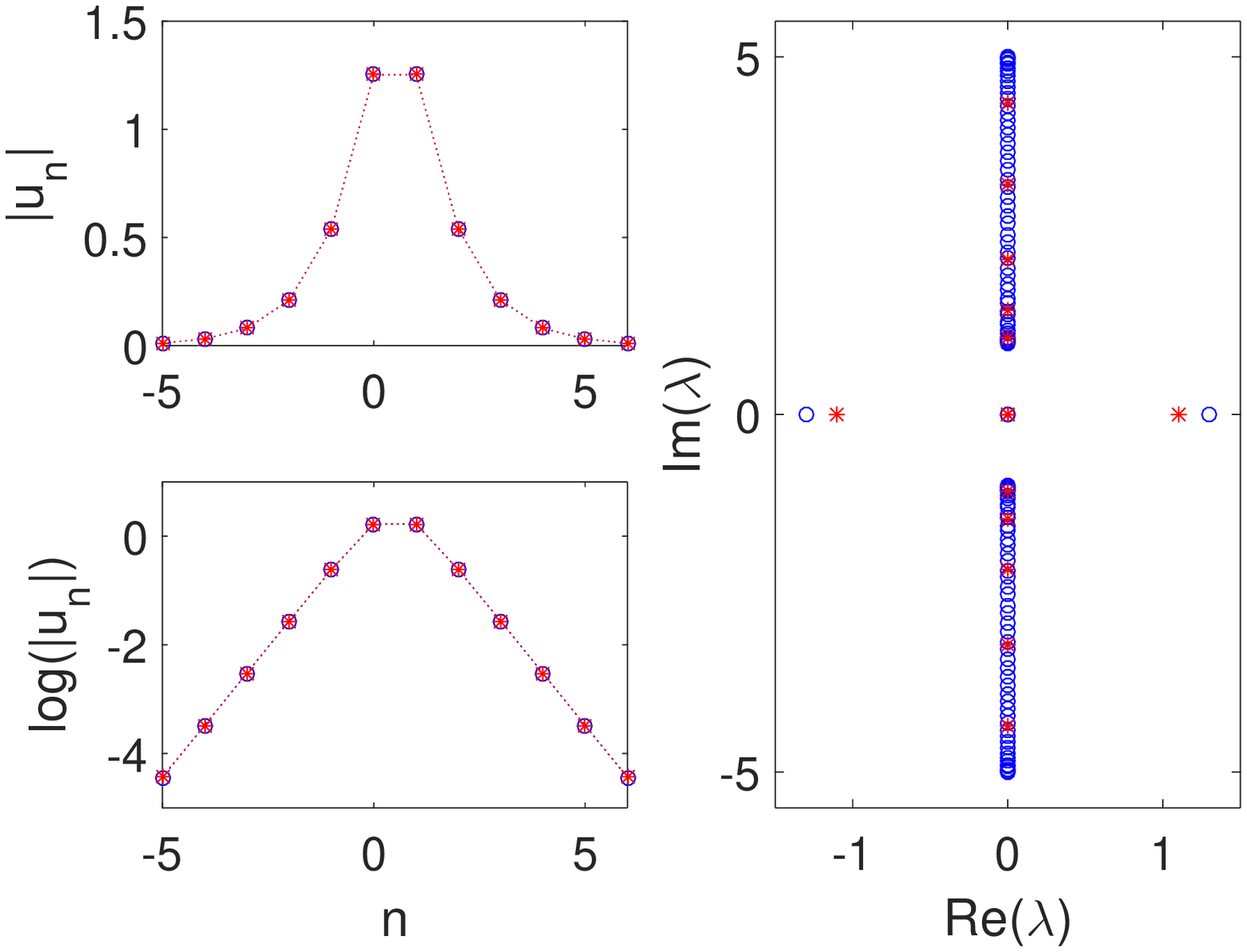}\label{subfig:intersite_N_3_c_1}}
	\caption{Comparison between the numerically obtained on-site (a,b)  and inter-site (c,d) solutions of \eqref{eq1} and Gaussian VA \eqref{ansatz2} and \eqref{sys} for $\omega=1, c=1$ and (a, c) $N=2$, (b, d) $N=3$. Corresponding spectrum of the solution profile is in the right panel of each figure. Blue circle-dashed and red star-dashed lines indicate numerical and Gaussian VA results, respectively.
	}
	\label{fig:onsite_N_1_3}
\end{figure*}

Figure \ref{fig:onsite_N_1_3} depicts the comparison of on-site and intersite solutions obtained from the numeric of time-independent DNLS \eqref{eq1} and multiple Gaussian VA for $c=1$. The results shown are for $N=2$ and $N=3$, where we can see that the solution profile is getting closer to the numerical result. The results also show that by increasing the number of Gaussian functions used in the VA, the approximation provides a better capture to the soliton amplitude and tail. We also depict in Fig. \ref{fig:onsite_N_1_3} the corresponding spectrum of the on-site and inter-site solitons. We can see that for on-site cases, the false instability no longer exists.

\begin{figure}[h]
	\centering
\includegraphics[scale=0.4]{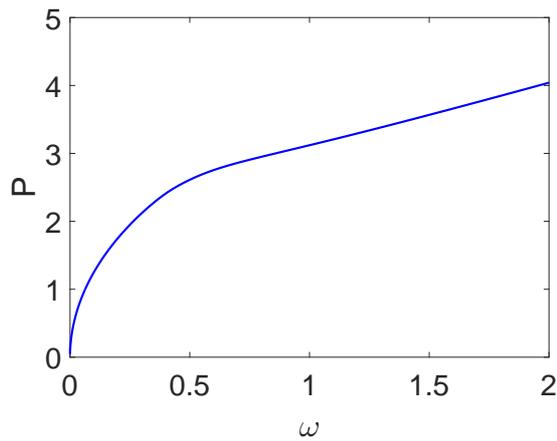} 
	\caption{Power of on-site solitons as  a function of $\omega$ for $c=1$ and $N=2$.
	}
		\label{fig:power_onsite_c_1_NN_2}
\end{figure}

We can also see the remedy of the false instability using Vakhitov-Kolokolov criterion for $N=2$. Figure \ref{fig:power_onsite_c_1_NN_2} shows the plot of $P$ for $c=1$ and $N=2$, where we can see that the on-site soliton is now stable for any value of $\omega$, i.e., there is no negative slope that existed in the case of $N=1$ shown in Fig.\ \ref{fig:power_onsite_c_1_NN_1}.

\section{Conclusions} \label{secconcl}
We have studied time independent solutions of DNLS equation and their stability by using VA method. We have employed the ansatz containing single and multiple Gaussian functions. We have shown that the method can be used to approximate the solution of DNLS and to analyse their stability. Analytically, we approximate the solution for using a single Gaussian function, and found that for the on-site case, there is a false instability. A remedy has been provided by increasing the number of Gaussian functions used in the ansatz, confirming the fact that the instability is caused by the shape of our ansatz, which has not been reported before. 

\section*{Acknowledgements}
R.R. (Grant Ref.\ No:\ S-5405/LPDP.3/2015) and R.K. (Grant Ref.\ No:\ S-34/LPDP.3/2017) gratefully acknowledge financial support from Lembaga Pengelolaan Dana Pendidikan (Indonesia Endowment Fund for Education). The authors gratefully acknowledge the two anonymous referees for their careful reading.

\appendix

\section{Variational Approximation} \label{secVA}
Following \cite{dawes2013variational}, consider a discrete differential equation of the form
\begin{equation} \label{pde}
i\dot{u}_n=f_V(u_n,u_n^*)+f_{NV}(u_n,u_n^*),
\end{equation} 
where $u_n(t)$ is a complex function of time and $u_n^*$ is its conjugate. We write the right hand side of (\ref{pde}) into two parts: the variational part $f_V$ and the nonvariational part $f_{NV}$. If the equation is conservative, then $f_{NV}=0$, and there is a Lagrangian function
\begin{equation} \label{Lang}
\mathcal{L}=\sum_{n=-\infty}^{\infty} L(u_n,\dot{u}_n, u_n^*,\dot{u}_n^*),
\end{equation}
that yields equation (\ref{pde}) from the relation
\begin{equation}
\frac{\partial \mathcal{L}}{\partial u_n^*}-\frac{d}{dt}\left(\frac{\partial \mathcal{L}}{\partial \dot{u}_n^*}\right)=i\dot{u}_n-f_V(u_n,u_n^*).
\end{equation}

Let $U_n$ be the chosen ansatz containing a finite number of parameters to be determined, $x_j$, for $j=1, 2, \cdots, N$, and they are functions of $t$. The calculation upon the substitution of the ansatz to the infinite sum (\ref{Lang}) yields an effective Lagrangian $\mathcal{L}_{eff}$, which contains the variational parameters and their derivatives with respect to $t$. This gives us the variational equations
\begin{equation} \label{ve1}
\frac{\partial \mathcal{L}_{eff}}{\partial x_j}-\frac{d}{dt}\left(\frac{\partial \mathcal{L}_{eff}}{\partial \dot{x}_j}\right)=0,
\end{equation}
which can be solved for $x_j$'s. 

Now, we consider the case when $f_{NV} \neq 0$. Recall that, 
\begin{eqnarray*}
	\frac{\partial \mathcal{L}_{eff}}{\partial x_j}&=&\frac{\partial \mathcal{L}_{eff}}{\partial U_n^*}\frac{\partial U_n^*}{\partial x_j}+\frac{\partial \mathcal{L}_{eff}}{\partial U_n}\frac{\partial U_n}{\partial x_j}\\
	&=& 2 \text{Re}\left(\sum_{n=-\infty}^{\infty}\left(i\dot{U}_n-f_V+\frac{d}{dt}\left(\frac{\partial L}{\partial \dot{U}_n^*}\right)\right) \frac{\partial U_n^*}{\partial x_j}\right).
\end{eqnarray*}
On the other hand, since the ansatz $U_n$ is assumed to satisfy equation (\ref{pde}), then
\begin{eqnarray*}
	\frac{\partial \mathcal{L}_{eff}}{\partial x_j}=2 \text{Re} \left(\sum_{n=-\infty}^{\infty} \left(f_{NV} +\frac{d}{dt}\left(\frac{\partial L}{\partial \dot{U}_n^*}\right)\right) \frac{\partial U_n^*}{\partial x_j}\right).
\end{eqnarray*}
Note that in the two last equations, $L$ is restricted to the space where the ansatz is defined. Combining them, we obtain the equation
\begin{equation} \label{ve2}
\text{Re}\left(\sum_{n=-\infty}^{\infty}i\dot{U}_n \frac{\partial U_n^*}{\partial x_j}\right)=\text{Re}\left(\sum_{n=-\infty}^{\infty}(f_V+f_{NV}) \frac{\partial U_n^*}{\partial x_j}\right).
\end{equation}
The equivalence of equation \eqref{ve2} for spatially continuous and linear Schr\"odinger equations is known as the Dirac-Frenkel-MacLachlan variational principle \cite{mclachlan1964variational, heller1976time} (see also \cite{lubich2005variational,faou2006poisson} that give a near-optimality result for variational approximations in that case, by providing error bounds in terms of the distance of the exact wave function to the approximation manifold).

\section{}
\label{secapp1}
In this section, we write equation \eqref{ODE} explicitly. Writing
\begin{align*}
	\varphi_s&=\sum_{n=-\infty}^{\infty}(n-n_0)^se^{-2a(n-n_0)^2},\\
	\phi_s&=\sum_{n=-\infty}^{\infty}(n-n_0)^se^{-4a(n-n_0)^2},\\
	\chi_s&=\sum_{n=-\infty}^{\infty}(n-n_0)^s\left(e^{-i\beta}e^{-2a(n-n_0)-i\gamma (n-n_0)-2a(n-n_0)^2}+e^{i\beta}e^{2a(n-n_0)+i\gamma_1 (n-n_0)-2a(n-n_0)^2}\right),
\end{align*}
the matrix $M$ is given by
\[{\small M=
\begin{pmatrix}
0&0&-A\varphi_0&-A\varphi_1&-\frac{1}{2}A\varphi_2&A(\beta\varphi_0+\gamma\varphi_1)\\
0&0&A^2\varphi_2&A^2\varphi_3&\frac{1}{2}A^2\varphi_4&-A^2(\beta\varphi_2+\gamma\varphi_3)\\
A \varphi_0&-A^2\varphi_2&0&0&0&2A^2a\varphi_1\\
0&-A^2\varphi_3&0&0&0&2aA^2\varphi_2\\
\frac{1}{2}A\varphi_2&-\frac{1}{2}A^2\varphi_4&0&0&0&A^2a\varphi_3\\
-A(\beta\varphi_0+\gamma\varphi_1)&A^2(\beta\varphi_2+\gamma\varphi_3)&-2A^2a\varphi_1&-2A^2a\varphi_2&-A^2a\varphi_3&0
\end{pmatrix},}\]
while the vector function $F$ is
\begin{equation*}
{\small F=
\begin{pmatrix}
A^3\phi_0+Ac\chi_0-A(\omega+2c) \varphi_0\\
-A^4\phi_2-A^2c\chi_2+A^2(\omega+2c)\varphi_2\\
\text{Re}(-iA^2ce^{-a+\frac{i\gamma}{2}}\chi_0)\\
\text{Re}(-iA^2ce^{-a+\frac{i\gamma}{2}}\chi_1)\\
\text{Re}\left(-\frac{1}{2}iA^2ce^{-a+\frac{i\gamma}{2}}\chi_2\right)\\
2A^2a(A^2\phi_1-(\omega+2c)\varphi_1)+\text{Re}(iA^2c\beta_1e^{-a+\frac{i\gamma}{2}}\chi_0+A^2(2a+i\gamma)ce^{-a+\frac{i\gamma}{2}}\chi_1)
\end{pmatrix}.}
\end{equation*}

\section{}
\label{secapp2}
By defining
\begin{eqnarray*}
	\psi_s&=&\sum_{n=-\infty}^{\infty}(n-n_0)^s\left(e^{-2a_1(n-n_0)-2a_1(n-n_0)^2}-e^{2a(n-n_0)-2a(n-n_0)^2}\right),
\end{eqnarray*}
we obtain the matrix component of $B$ in \eqref{EVP} 

\begin{align*}
	{\small \begin{aligned}
		b_{11}&=3A^2\phi_0-(\omega+2c)\varphi_0+ce^{-a}\chi_0, & b_{12}&=-Ace^{-a}\chi_0-2Ace^{-a}\psi_1-4A^3\phi_2 \\ & & & \quad +2A(\omega+2c)\varphi_2-2Ace^{-a}\chi_2\\
		b_{13}&=0, & b_{14}&=0\\
		b_{15}&=0, &  b_{16}&=2aAce^{-a}\psi_0\\
		b_{21}&=-4A^3\phi_2+2A(\omega+2c)\varphi_2 & b_{22}&=A^2ce^{-a}\chi_2+2A^2ce^{-a}\psi_3+4A^4\phi_4 \\
		&\quad -2Ace^{-a}\chi_2, & & \quad -2A^2(\omega+2c)\varphi_4+2A^2ce^{-a}\chi_4\\
		b_{23}&=0,  & b_{24}&=0\\
		b_{25}&=0, &  b_{26}&=-2aA^2ce^{-a}\psi_2-4aA^2ce^{-a}\chi_3\\
		b_{31}&=0, &  b_{32}&=0\\
		b_{33}&=0, &  b_{34}&=A^2ce^{-a}\psi_0 \\
		b_{35}&=\frac{1}{2}A^2ce^{-a}\chi_0+A^2c\psi_1, &  b_{36}&=0 \\
		b_{41}&=0, &  b_{42}&=0 \\
		b_{43}&=0, &  b_{44}&=A^2ce^{-a}\psi_1 \\
		b_{45}&=A^2ce^{-a}\psi_2, &  b_{46}&=0\\
		b_{51}&=0, &  b_{52}&=0\\
		b_{53}&=0, &  b_{54}&=\frac{1}{2}A^2ce^{-a}\psi_2 \\
		b_{55}&=\frac{1}{4}A^2ce^{-a}\chi_2+\frac{1}{2}A^2ce^{-a}\psi_3, & b_{56}&=0 \\
		b_{61}&=0, &  b_{62}&=-4aA^2ce^{-a}\psi_2-4aA^2ce^{-a}\chi_3 \\
		b_{63}&=0, &  b_{64}&=0\\
		b_{65}&=0, &  b_{66}&= -2aA^4\phi_0+2aA^2(\omega+2c)\varphi_0-2aA^2ce^{-a_1}\chi_0 \\
		& &  & \quad +16a^2A^4\phi_2 -8a^2A^2(\omega+2c)\varphi_2\\
		& &  & \quad +4a^2A^2ce^{-a}(2\chi_2+\psi_1).
	\end{aligned}}
\end{align*}

\section*{References}









\end{document}